\documentclass[fleqn,usenatbib]{mnras}

\usepackage{newtxtext,newtxmath}

\usepackage[T1]{fontenc}

\DeclareRobustCommand{\VAN}[3]{#2}
\let\VANthebibliography\thebibliography
\def\thebibliography{\DeclareRobustCommand{\VAN}[3]{##3}\VANthebibliography}

\usepackage{graphicx}	
\usepackage{amsmath}	
\usepackage{amssymb}	
\usepackage{booktabs}   
\usepackage{multirow}   
\usepackage{comment}    
\usepackage[utf8]{inputenc} 

\title[Insights into galaxy quenching]{
Different regulation of stellar metallicities between star-forming and quiescent galaxies -- Insights into galaxy quenching}
\author[W. M. Baker et al.]
{William M. Baker$^{1,2}$\thanks{E-mail: wb308@cam.ac.uk},
Roberto Maiolino$^{1,2,3}$,
Asa F. L. Bluck $^{4}$,
Francesco Belfiore $^{5}$,
Mirko Curti $^{1,2,6}$,
\newauthor{}
Francesco D'Eugenio $^{1,2}$,
Joanna M. Piotrowska $^{1,2,7}$,
Sandro Tacchella $^{1,2,}$,
James A. A. Trussler $^{8}$\\
$^{1}$Kavli Institute for Cosmology, University of Cambridge, Madingley Road, Cambridge, CB3 OHA, UK\\
$^{2}$Cavendish Laboratory - Astrophysics Group, University of Cambridge, 19 JJ Thomson Avenue, Cambridge, CB3 OHE, UK\\
$^{3}$Department of Physics and Astronomy, University College London, Gower Street, London WC1E 6BT, UK\\
$^{4}$Department of Physics, Florida International University, 11200 SW 8th Street, Miami, FL, USA\\
$^{5}$INAF— Osservatorio Astrofisico di Arcetri, Largo E. Fermi 5, I-50125, Florence, Italy\\
$^{6}$European Southern Observatory, Karl-Schwarzschild-Strasse 2, D-85748 Garching bei Muenchen, Germany\\
$^{7}$Cahill Center for Astronomy and Astrophysics, California Institute of Technology, Pasadena, CA 91125, USA\\
$^{8}$Jodrell Bank Centre for Astrophysics, University of Manchester, Oxford Road, Manchester M13 9PL, UK\\
}

\date{Accepted XXX. Received YYY; in original form ZZZ}

\pubyear{2021}

\begin{document}
\label{firstpage}
\pagerange{\pageref{firstpage}--\pageref{lastpage}}
\maketitle

\begin{abstract}
One of the most important questions in astrophysics is what causes galaxies to stop forming stars. Previous studies have shown a tight link between quiescence and black hole mass. Other studies have revealed that quiescence is also associated with `starvation', the halting of gas inflows, which results in the remaining gas being used up by star formation and in rapid chemical enrichment. In this work, we find the missing link between these two findings. Using a large sample of galaxies, we uncover the intrinsic dependencies of the stellar metallicity on galaxy properties. In the case of star-forming galaxies, stellar metallicity is primarily driven by stellar mass. However, for passive galaxies, the stellar metallicity is primarily driven by the stellar velocity dispersion. The latter is known to be tightly correlated with black hole mass. This result can be seen as connecting previous studies, where the integrated effect of black hole feedback (i.e. black hole mass, traced by the velocity dispersion) prevents gas inflows, starving the galaxy, which is seen by the rapid increase in the stellar metallicity, and leading to the galaxy becoming passive. 
\end{abstract}




\section{Introduction}

The chemical enrichment of galaxies is a powerful proxy of their history of star formation and of the evolutionary processes and mechanisms that have shaped them. This chemical enrichment is traced by the `metallicity', the ratio of heavier elements to the Hydrogen and Helium content of either gas or stars. Metals are produced by stellar nucleosynthesis which are then released via supernovae explosions and stellar winds, hence increasing the metallicity of the ISM \citep{Maiolino2019}. However, metallicity also traces other processes, such as inflows of low-metallicity gas (causing metal dilution) and outflows \citep[often metal loaded,][]{Origlia2004,
Chisholm2018}{}{}.

Depending on the galaxy type and on the available data, the metallicity in galaxies can be measured for the gas-phase of the ISM, and for the stellar population. Both of these metallicities follow well-known scaling relations with other galactic properties. Specifically, metallicity is known to increase with the stellar mass of a galaxy, up until $\rm M_*=10^{10.5} M_\odot$ where it starts to plateau, in a relationship known as the mass-metallicity relation \citep[MZR, ][]{Tremonti2004, Zahid2017, Maiolino2019}{}{}. At fixed mass the metallicity is also observed to decrease with increasing star-formation rate (i.e. anti-correlate), leading to an overall three-dimensional relationship between the quantities, known as the Fundamental Metallicity Relation \citep[FMR,][]{Mannucci2010, Baker2023b,Looser23,Ellison2008}{}{}.

 Other studies have found correlations between stellar metallicity and properties such as stellar velocity dispersion and stellar age \citep{Li2018, Cappellari2023}, but have been unable to probe multiple quantities simultaneously; hence, they have been unable to disentangle whether these correlations are intrinsic, or indirect by-products of other quantities.

The advantage of studying stellar metallicities is that they can be estimated for both star-forming and passive galaxies, whereas, due to being calculated from emission lines, gas-phase metallicities can only accurately be obtained for star-forming galaxies \citep[although see also ][]{Kumari2019}{}{}.
Star-forming galaxies and passive galaxies show large differences in stellar metallicity, with passive galaxies having larger values at a given stellar mass \citep{Peng2015,
Trussler2020,Trussler2021}. This is also seen as a continuous sequence in \cite{Looser23}, where galaxies with the lower SFR (i.e. more passive) have higher stellar metallicity. This trend is interpreted as quenching via starvation, where galaxies quench because their gas supply (via inflows) is cut off \citep{Peng2015, Trussler2020,Bluck2020b}. Starvation implies that the remaining gas of the galaxy is converted into stars in a closed-box manner, resulting in the metallicity rising steeply because it is not diluted from accreting (low-metallicity) gas. However, even if starvation is identified as playing a key role in galaxy quenching, the causes of starvation and the associated quenching could not be identified based solely on the metallicity analysis.

Many different causes for the quenching of star-formation have been proposed in the past, such as AGN feedback, morphological quenching, cosmic rays, magnetic fields, and environmental effects \citep{Man2018}. Of particular interest are the recent results of \cite{Piotrowska2022} and \cite{Bluck2022, Bluck2023}, where they find that the strongest link to quenching in a galaxy is the mass of the supermassive black hole at the centre (this applies for central galaxies or massive satellites). This has led \cite{Bluck2022} to conclude that it is the integrated effect of black hole accretion feedback that causes quenching. The proposed scenario, also supported by simulations \citep{Piotrowska2022,Zinger2020}, is that black hole accretion heats the circum-galactic medium (via radio-jets or outflows) hence preventing accretion of cold gas and resulting in delayed quenching via starvation.

A key difficulty when investigating relationships between quantities that could cause quenching is that many of them are inter-correlated, hence correlation between quiescence and some galaxy properties does not necessarily imply causation. In the recent works discussed above \citep{Piotrowska2022, Bluck2022, Bluck2023,Bluck2024}, this issue was tackled by using machine learning (random forest) and partial correlation techniques to disentangle direct and primary connections from indirect relations induced by spurious connections resulting from the fact that most galactic properties are intercorrelated. The same issue applies also for the metallicity scaling relations: it is hard to separate intrinsic fundamental relationships from secondary relations resulting from indirect byproducts. Within this context \cite{Baker2022, Baker2023a, Baker2023b} and \cite{Baker2023c} utilised multiple techniques to break these inter-correlations and determine the fundamental dependencies of gas-phase metallicities and other scaling relations.

In this paper we extend those studies to explore the intrinsic dependence of {\it stellar} metallicity as a function of galactic properties, finding the connection between the mechanism of quenching, starvation, and the likely cause, integrated black hole feedback.

In this paper, we assume that $\rm H_0=70$km $\text{s}^{-1}$ $\text{Mpc}^{-1}$, $\Omega_m=0.3$ and $\Omega_{\Lambda}=0.7$.

\section{Data and samples}
\label{sec:Data_and_methods}

\begin{figure*}
    \centering
    \includegraphics[width=1\columnwidth]{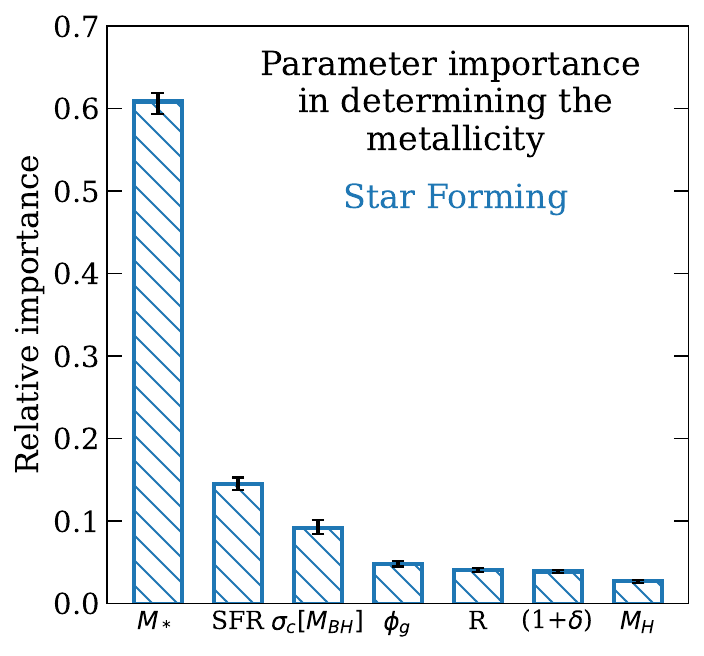}
    \includegraphics[width=1\columnwidth]{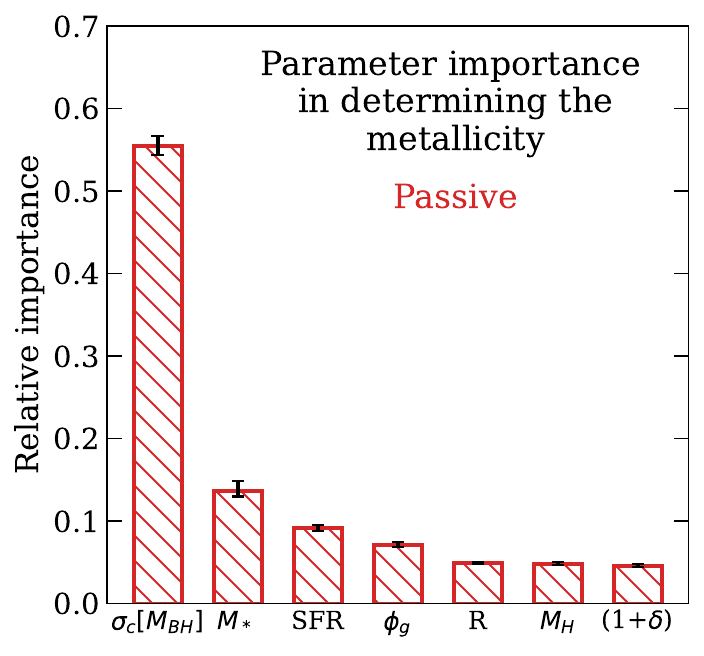}
    \caption{ The importance of various galactic properties in determining the stellar metallicity (Z) for star-forming (left panel) and passive (right panel) galaxies. The galactic parameters evaluated are: the stellar mass ($M_*$), the star formation rate (SFR), the central stellar velocity dispersion ($\sigma_c$), the gravitational potential (traced by $\phi_g = M_*/r_e$), a control uniform random variable (R), the overdensity of galaxies (1+$\delta$), and the halo mass ($M_H$). The left hand panel shows that for star-forming galaxies the primary dependence of stellar metallicity is on stellar mass via the mass-metallicity relation (MZR). However, for passive galaxies (right hand panel) the stellar metallicity dependence is almost purely on  central velocity dispersion, which is know to tightly trace the black hole mass.}
    \label{fig:RF}
\end{figure*}

\subsection{SDSS Data}
\label{sec:data}

We use data from the Sloan Digital Sky Survey (SDSS) data release 7 \citep{SDSS_DR7_2009ApJS..182..543A} MPA-JHU catalogue \citep{Brinchmann2004, Tremonti2004}. We then match this catalogue to another used in \cite{Trussler2020}, containing mass and light weighted stellar metallicities fit by the spectral fitting code FireFly \citep{FIREFLY_2017MNRAS.472.4297W}. FIREFLY is a spectral fitting code that uses $\chi^2$ minimisation using a linear combination of simple stellar populations which vary in stellar ages and metallicities. The stellar population models used are those of \cite{Maraston2011} as well as input spectra from MILES \citep{MILES_2006MNRAS.371..703S} and a Kroupa IMF \citep{Kroupa2001}.
We also use mass-weighted halo masses and central satellite classifications from \cite{Yang2005, Yang2007}, and overdensities from \cite{Peng2010}. We also match to a catalogue of central velocity dispersions from the NYU value added catalogue \citep{Blanton2005} and morphological parameters from \cite{Simard2011}. 
We sigma clip the resulting dataset to 3$\sigma$ to remove outliers that could bias the random forest or partial correlation coefficients. We show the resulting sample for SDSS in Table \ref{table:samples}.
This dataset enables us to explore the drivers of the stellar metallicity for a large number of galaxies (15,000+ in each sample) whilst also being able to explore many parameters of interest relating to each individual galaxy and its environment.

\begin{table}
\caption{Number of galaxies contained within each SDSS sample once all cuts and matches have been made. Samples are star-forming stellar metallicities (SF Z sample) and non-star-forming stellar metallicities (NSF Z sample).  Can see that there are many more passive galaxies than SF, hence the need to separate out these classes to properly explore the drivers. }
\centering
\label{table:samples}
\begin{tabular}{l c c c c c}
\toprule
\multirow{2}{*}{} &  \multirow{2}{*}{\# of galaxies} & \multirow{2}{*}{log($M_*/M_\odot$)} & \multirow{2}{*}{log(SFR$/M_\odot$yr)} & \multirow{2}{*}{Overdensity} \\[+5pt]
\midrule
SF & 16505 & 9.0 - 11.6 & -1.0 - 1.5 & -1.2 - 2.1 \\
Q & 37241 & 9.6 - 11.7 & -2.1 - 0 & -1.2 - 2.7 \\
\bottomrule
\end{tabular}
\end{table}

\subsection{MaNGA Data}
\label{sec:manga}
We also use data from the Mapping nearby Galaxies at Apache Point Observatory (MaNGA) Survey \citep{Bundy2015}, which enables us to also include accurately measured dynamical masses from \cite{Li2018} and kinematic measurements from \cite{Brownson2022}. We use stellar metallicities, stellar masses, central stellar velocity dispersions, and star-formation rates from \cite{Pipe3d_1_2016RMxAA..52...21S, Pipe3d_2_2016RMxAA..52..171S}.
This significantly cuts down our sample statistics to around 1000 galaxies each for the star-forming and passive samples, but enables us to control for these quantities which were unavailable in the larger SDSS sample.

\subsection{Definition of the sample}

Regarding the SDSS data, in order to ensure reliable stellar metallicity measurements, we employ a strict median signal to noise cut of 20 per spectral pixel \citep[as in][]{Trussler2020}. We also limit the sample to those galaxies with reliable redshifts in the range $0.02\leq z \leq 0.085$ to both reduce aperture effects (i.e. the projected size of the SDSS fibre should be larger than 1~kpc) and reduce the effects of cosmological evolution \citep[again, as in][]{Trussler2020}.

In both the SDSS and MaNGA samples we divide our sample of galaxies into star-forming and passive based on their distance from the main sequence \citep[using the MS definition of][]{Renzini2015}{}{}. Star-forming galaxies are those with $\rm \Delta MS\geq-0.75$, whilst passive galaxies are those with $\rm \Delta MS\leq-0.75$.
This gives us 16505 star-forming galaxies and 37241 passive galaxies.

\subsection{Velocity dispersion and black hole masses}

 We calculate the central velocity dispersion using the approach of \cite{Jorgensen1996} and \cite{Bluck2016}. To briefly summarise, we use the equation
\begin{equation}
    \rm \sigma_c=\sigma_v \times \bigg(\frac{R_{bulge}}{8\times R_{ap}}\bigg)^{-0.04} 
\end{equation}
 where $\sigma_c$ is the central (stellar) velocity dispersion, $\sigma_v$ is the stellar velocity dispersion, $\rm R_{bulge}$ is the bulge radius, and $\rm R_{ap}$ is the aperture radius (for SDSS this corresponds to the 1.5'' fibre). 

As pointed out by multiple works, the stellar velocity dispersion is the quantity most tightly correlating with the black hole mass \citep[see review in ][]{Kormendy2013}. So, when exploring the dependence on stellar velocity dispersion we also show the corresponding black hole mass calculated via the equation from \cite{Saglia2016}, which is given by
\begin{equation}
    \rm M_{BH}=5.246 \times log\; \sigma_c\; -3.77.
\end{equation}

\section{Data analysis tools}

\subsection{Random Forest Regression}
\label{sec:RF}

A random forest is a collection of many different decision trees, each of which attempts to reduce a quantity called mean squared error (MSE, which itself measures the quality of each split) and which enables it to calculate parameters importance in driving a target variable. 
One of the key benefits of random forest regression is that it can probe several intercorrelated parameters simultaneously (allowing us to investigate many parameters at once). It can also find non-monotonic relationships between them. Furthermore, random forests have previously been found to be able to find intrinsic dependencies among many intercorrelated variables \citep[provided the quantity the target truly depends on is included,][]{Bluck2022}. They have also been shown to be able to successfully reverse engineer simulations in \cite{Bluck2022} and \cite{Piotrowska2022}. We include the random forest hyper-parameters and the passive galaxy sample mean squared errors of the test and training sample in Table \ref{tab:rf_params}.

\subsection{Partial correlation coefficients}
\label{sec:pcc}
Partial correlation coefficients allow us to take the correlation between two quantities A and B whilst controlling for any further correlation with quantity C. This means that if A and C are truly correlated, and B has a correlation with C this correlation will not introduce an unphysical observed correlation between A and B. 
This is important to determine whether the relationships are intrinsic or stem from relationships between other quantities that are not considered.
We use partial correlation coefficients to determine the direction of any found relationship, i.e. as the random forest only gives us their importances, the PCCs can tell us whether two quantities correlate or anti-correlate. In addition to this, it can provide another, although less powerful, method for visually checking the random forest results.
The ratios of partial correlation coefficients can be used to define a gradient arrow on a 3D diagram, or a 2D diagram in which the third variable is colour coded \citep{Bluck2020, Baker2022}
\begin{equation}
    \rm \text{tan}(\theta)=\frac{\rho_{yz|x}}{\rho_{xz|y}}
    \label{theta}
\end{equation}
where $\theta$ is the arrow angle measured anticlockwise from the horizontal while $\rm \frac{\rho_{yz|x}}{\rho_{xz|y}}$ is the ratio of the partial correlation coefficients between the y axis and colour-coded (z axis) quantities (whilst controlling for the x axis quantity), and between the x axis and colour-coded (z axis) quantities (whilst controlling for the y axis quantity). 
The gradient arrow then points in the direction of the greatest increase in the colour-coded (z-axis) quantity - it enables the dependence of that quantity on the x and y axis quantities to be identified visually, whilst the angle of the arrow enables it to be determined quantitatively. As an example, if the x and y axis quantities contributed equally to determining the colour-coded quantity, we would expect an arrow angle of 45 degrees.

\begin{figure*}
    \centering
    \includegraphics[width=0.98\columnwidth]{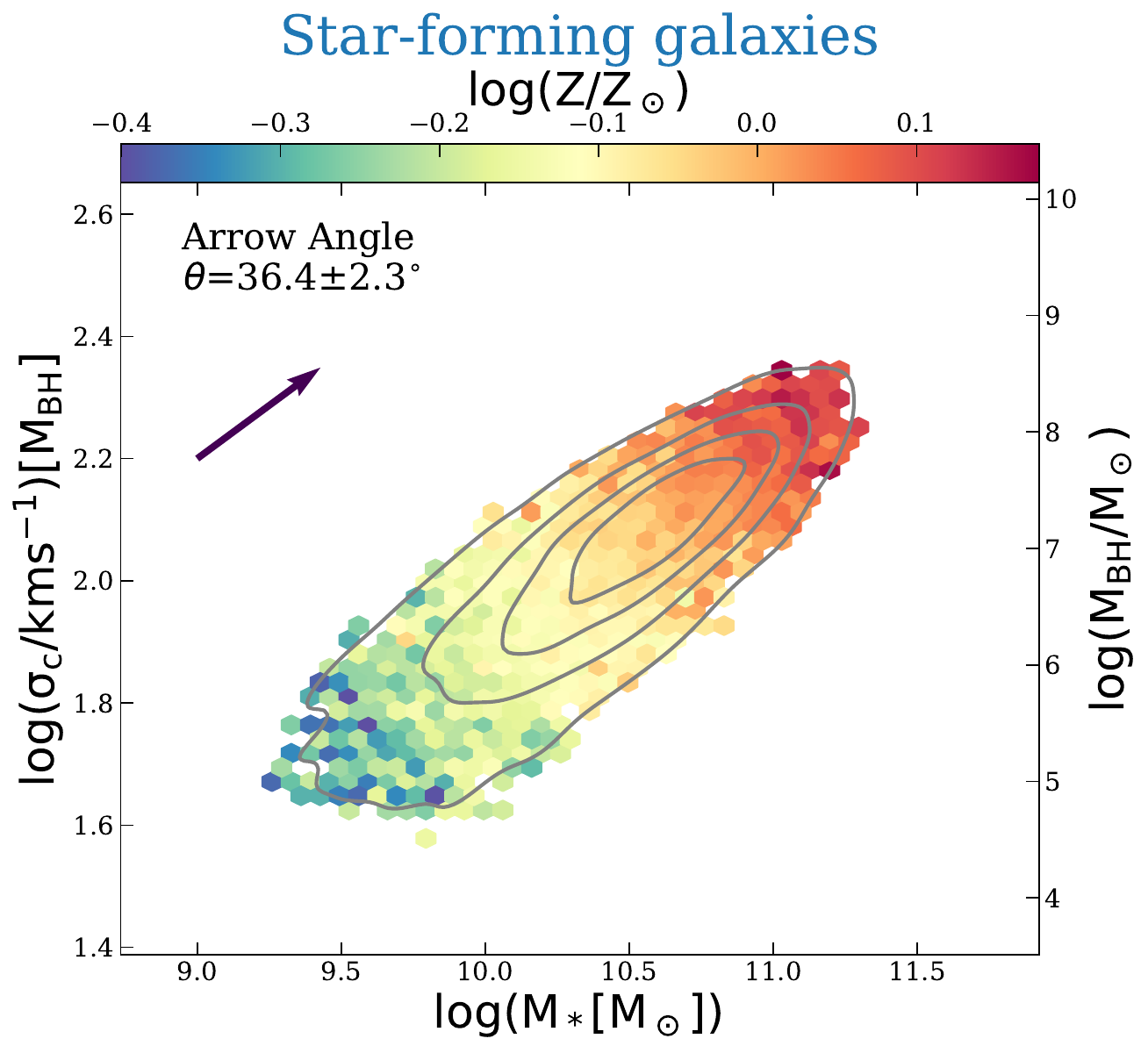}
    \includegraphics[width=1\columnwidth]{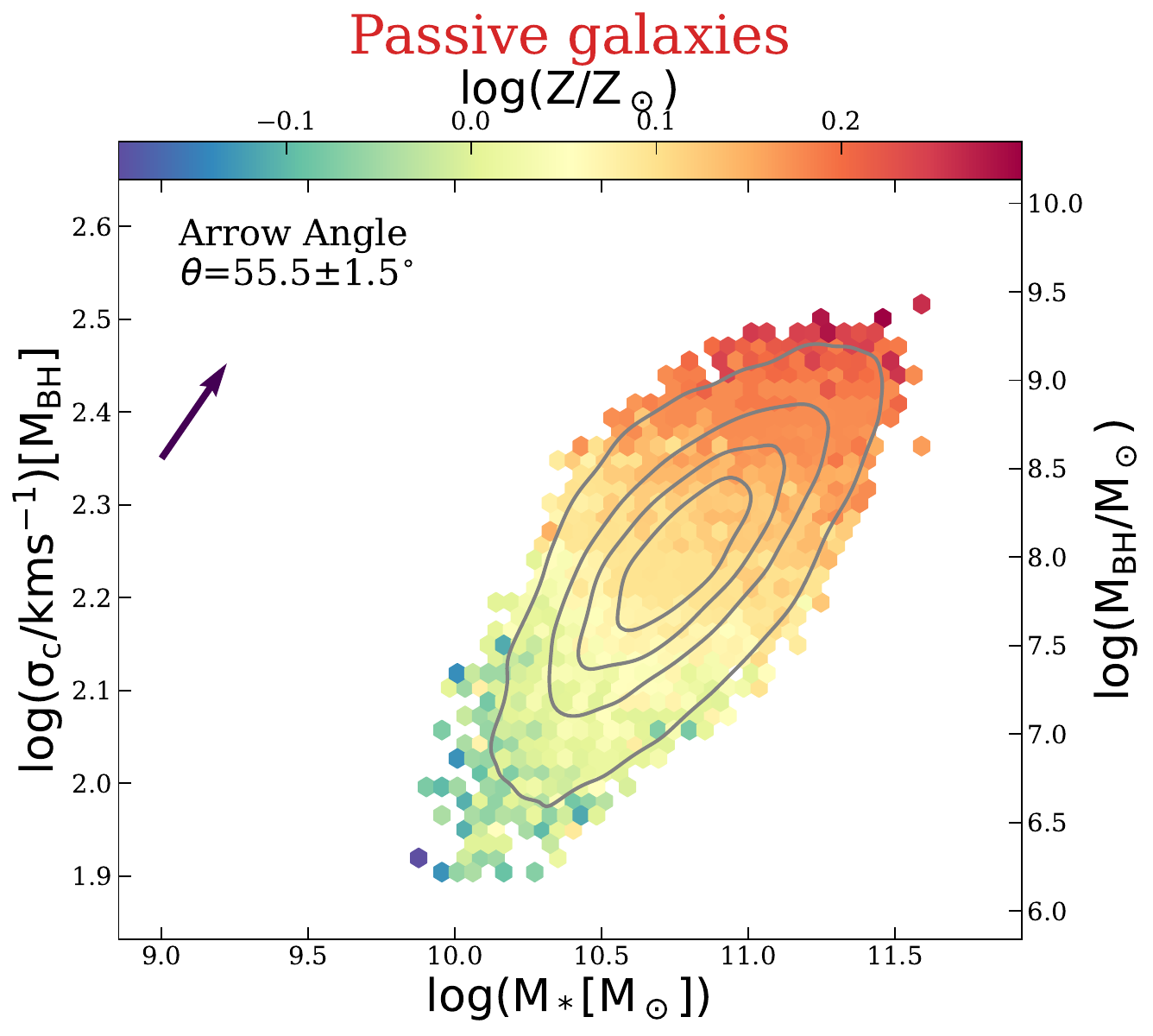}
    \caption{ Central stellar velocity dispersion versus stellar mass colour coded by median stellar metallicity for star-forming galaxies (left panel) and passive galaxies (right panel). The arrow angle points in the direction of the greatest increasing gradient of the stellar metallicity (i.e. it quantifies the colour-shading). The arrows and colour-shading gradients show that, in the left panel for star-forming galaxies, increasing stellar metallicity is linked to increasing stellar mass (i.e. the arrow angle is closer to the horizontal), whilst in the right panel for passive galaxies, the stellar metallicity is linked more to increasing  central velocity dispersion (the arrow angle is closer to vertical than horizontal). This indicates that the stellar metallicity of star-forming galaxies shows a stronger dependence on stellar mass than  central  velocity dispersion, whilst for passive galaxies this is reversed.  The right hand axis gives the corresponding black hole mass when assuming that it is traced by the central velocity dispersion.}
    \label{fig:hex}
\end{figure*}

\section{Results: what drives the stellar metallicity?}
\subsection{The key parameters according to Random Forest regression}

Our primary analysis tool is the random forest (RF) regression, as discussed in section 2.3. In this case the target feature is the stellar metallicity of a galaxy. The parameters that we explore are: stellar mass ($M_*$), star formation rate (SFR), the gravitational potential as approximately traced by $\Phi_g=M_*/R_e$ (where $R_e$ is the circularized half-light radius in the r-band), the mass of the dark matter halo ($M_H$) and the strength of the overdensity of galaxies (1+$\delta$). 
 
Among the parameters we also include the the central stellar velocity dispersion ($\sigma_c$). As already discussed,
various authors have shown that the central velocity dispersion is tightly linked to the central black hole mass \citep{Saglia2016, Piotrowska2022}, although we note that the central velocity dispersion might correlate with the dynamical mass or gravitational potential. However, later on we will discuss (and show) that neither dynamical mass nor gravitational potential appear play a role in determining the stellar metallicity.
Finally, we also include a uniform random variable as a control. 

We split the larger SDSS sample 50-50 into a training and test sample. The random forest is applied to the training sample to build a model which is then applied to the test sample to find the parameter importances. Errors for the parameter importances are calculated by bootstrap random sampling the random forest regression 100 times - we then take the 16th and 84th percentiles of the resulting distribution. We compare the test and training sample's mean squared errors (MSE) to ensure that the model is not overfitting. 
 For the random forest regression for the passive galaxies, we find a MSE of 0.007 for the testing sample and 0.005 for the training sample. For the star-forming galaxies, it is 0.017 and 0.013 respectively. This shows that there has been no significant overfitting or underfitting.

Fig. \ref{fig:RF} is a bar-chart showing the relative importance of the various quantities in determining the stellar metallicities, as inferred from the random forest analysis, for star-forming galaxies (left panel, blue) and passive galaxies (right panel, red). For the star-forming galaxies, the stellar metallicity is primarily driven by stellar mass (i.e. the MZR) with a secondary contribution from the SFR \citep[i.e. the FMR, as in ][]{Looser23}{}{}. However, for passive galaxies, the stellar metallicity is primarily driven by the central velocity dispersion with a small secondary importance on $M_*$ and SFR. Hence, the relative importance between stellar mass and central velocity dispersion (hence, black hole mass, if traced by $\sigma _c$)  {\it swap} depending on whether the galaxies are star-forming or passive.

One key aspect to note is that we see little to no dependence on $\Phi_g=M_*/R_e$, which is a proxy of the gravitational potential - this indicates the importance of $\sigma_c$ is not simply linked to the gravitational potential instead. Additionally, later on we use two smaller samples, but with spatially resolved spectroscopic information from the MaNGA survey \citep{Bundy2015}, for which the dynamical masses and gravitational potential could be accurately estimated. That analysis will show that neither the dynamical mass nor the gravitational potential of the galaxy drive the stellar metallicity, while also in that sample, central stellar velocity dispersion is the primary driver.

Finally, Fig.~\ref{fig:RF} shows that there is no significant dependence of the stellar metallicity on environment ($M_H$ or $(1+\delta)$), for either the star-forming or passive galaxies. In section \ref{sec:centvssat} we also verify that this result holds also when separating the sample between central and satellite galaxies.

\begin{figure*}
    \centering
    \includegraphics[width=1\columnwidth]{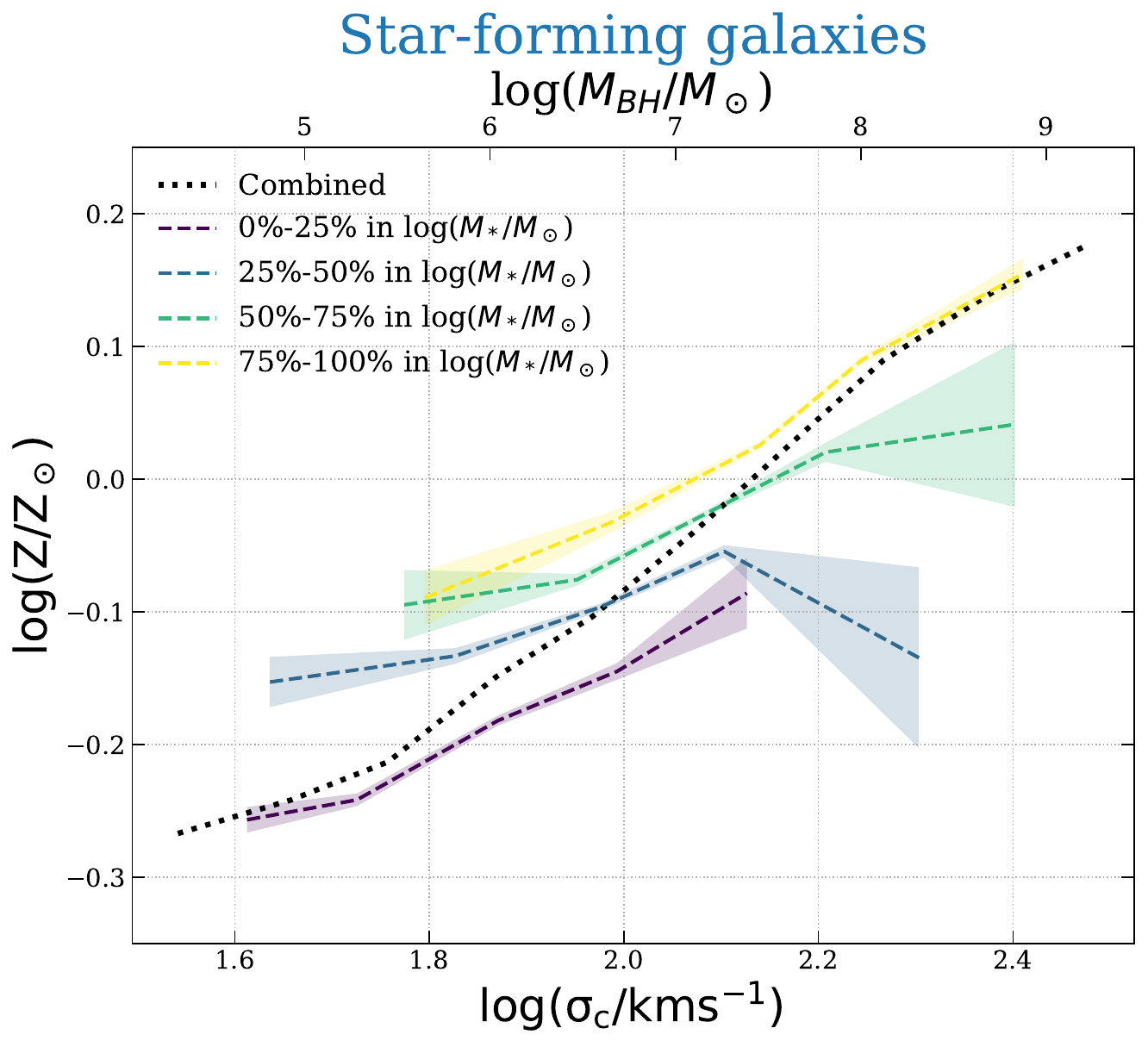}
    \includegraphics[width=1\columnwidth]{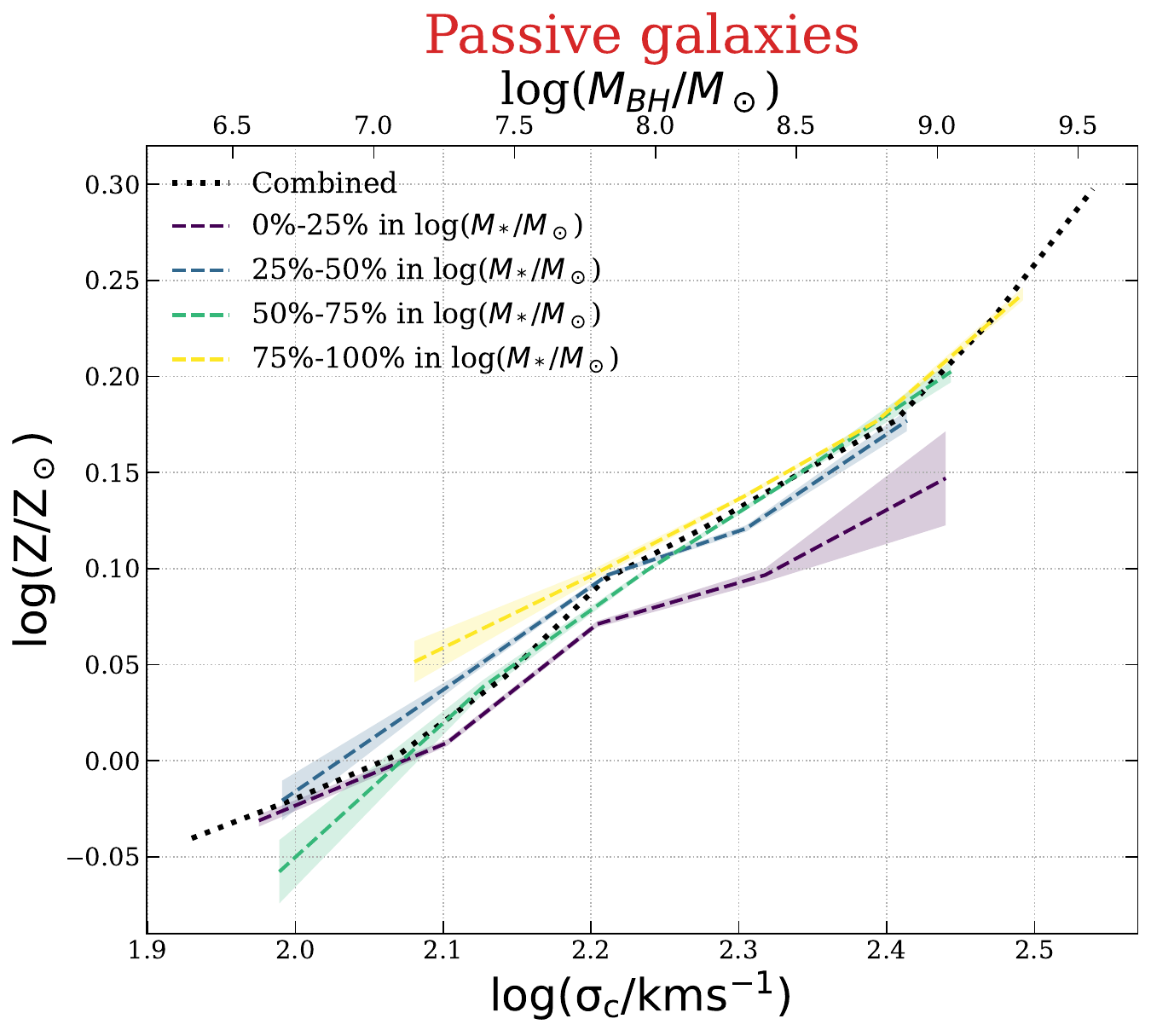}
    \includegraphics[width=1\columnwidth]{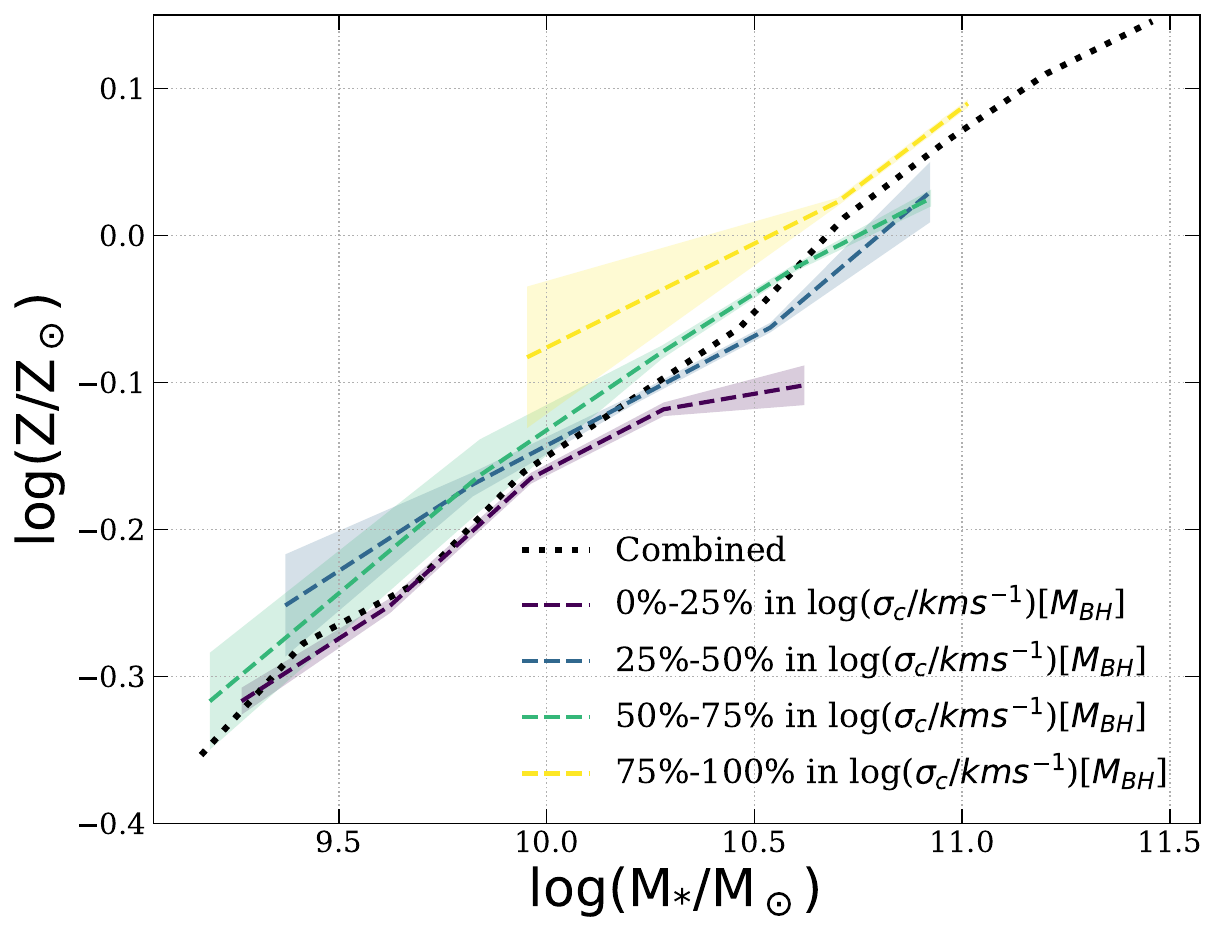}
    \includegraphics[width=1\columnwidth]{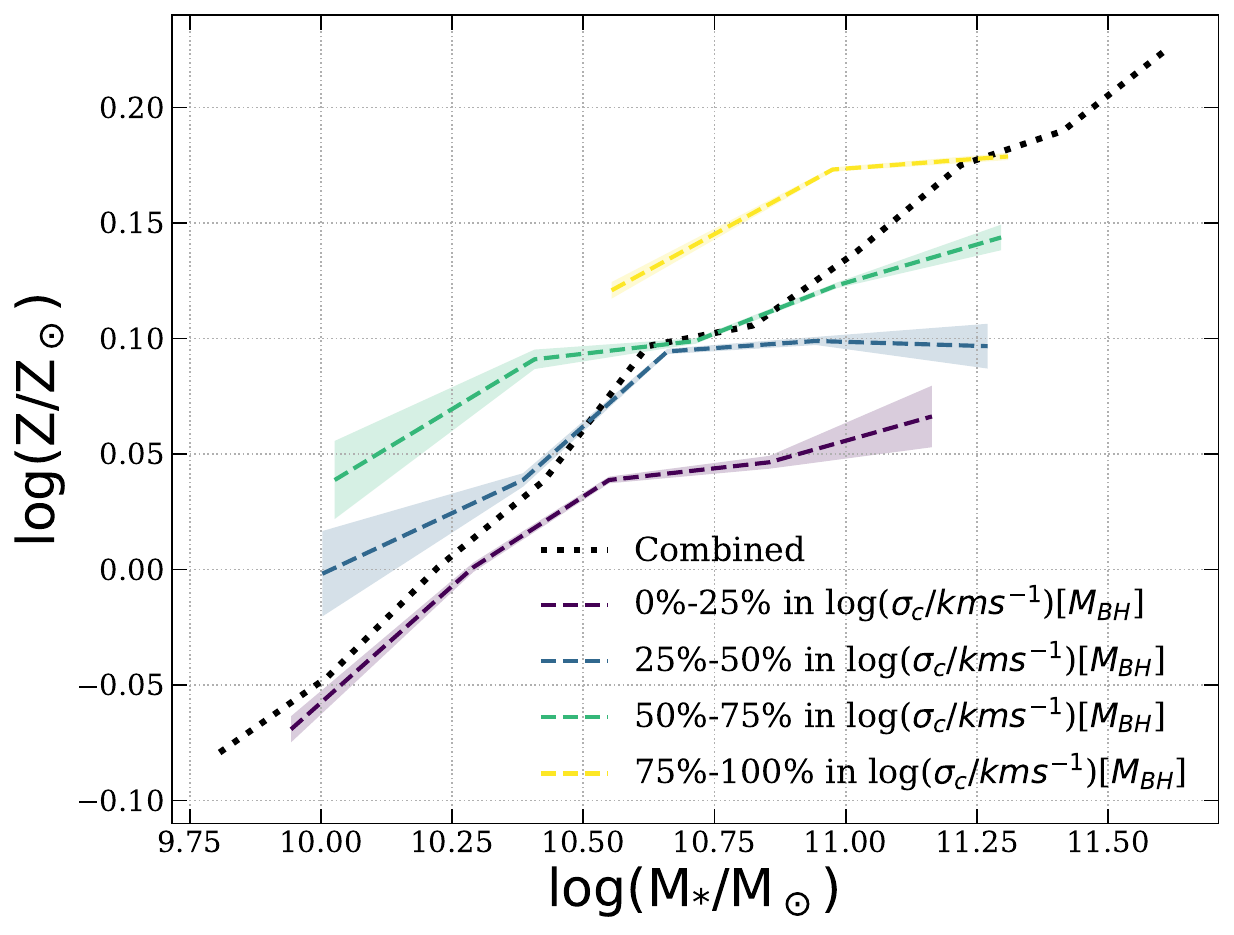}
    \caption{Upper: stellar metallicity versus  central stellar velocity dispersion binned by tracks of stellar mass, for star-forming galaxies (left, blue) and passive galaxies (right, red). Lower: stellar metallicity versus stellar mass binned by tracks of central velocity dispersion, again for star-forming (left, blue) and passive galaxies (right, red).
     The top axes give the corresponding black hole mass when assuming that it is traced by the central velocity dispersion.
    The left-hand plots show that the evolution in stellar metallicity for star-forming galaxies is primarily driven by stellar mass, whilst the right-hand plots show that for passive galaxies the stellar metallicity is primarily driven by central velocity dispersion (and, as we suggest in the text, by black hole mass.}
    \label{fig:tracks}
\end{figure*}

\begin{figure}
    \centering
    \includegraphics[width=\columnwidth]{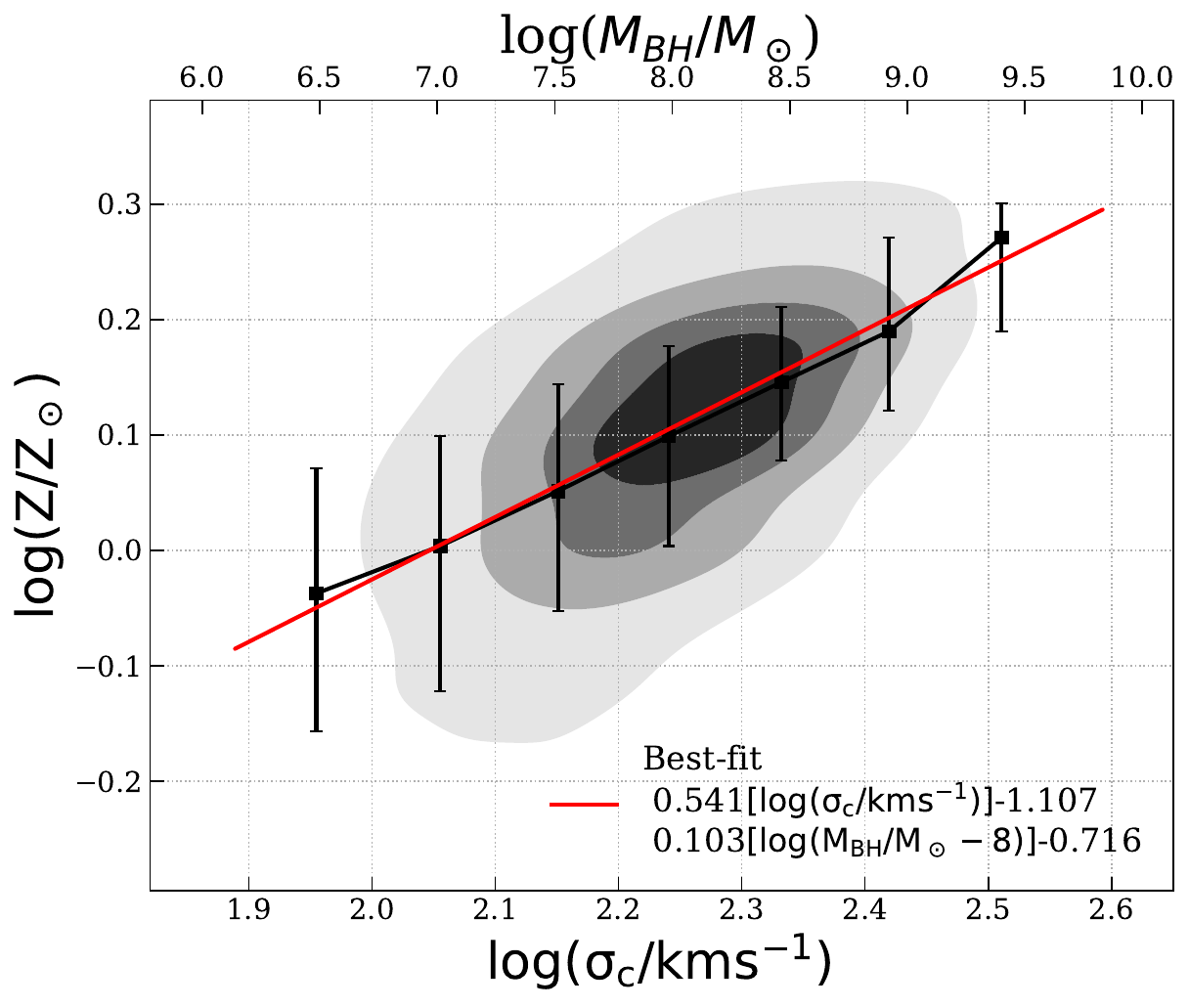}
    \caption{Stellar metallicity versus  central stellar velocity dispersion for passive galaxies.
      The top axis gives the corresponding black hole mass when assuming that it is traced by the central velocity dispersion. The outer density contour contains 90\% of the galaxy population. The black squares show the median values in bins of  central velocity dispersion with the errorbars corresponding to the 16th and 84th percentiles of the distributions in each bin.
    The red line is the best-fit to these bins,  showing the central velocity dispersion metallicity relation ($\sigma$ZR), or equivalent black hole mass metallicity relation BZR.}
    \label{fig:best_fit}
\end{figure}

\subsection{Direction of the relationship - partial correlation coefficients}
We next investigate more visually the observed dependence of stellar metallicity on stellar mass and central velocity dispersion in star-forming and passive galaxies, respectively, by reducing the dimensionality of the problem, specifically using `simple' 3D diagrams \citep{Bluck2020}, which also allow us to assess the sign of the relation, i.e. whether positive or negative, which is not possible via the random forest.

Figure \ref{fig:hex} shows the central velocity dispersion versus stellar mass, colour-coded by the median stellar metallicity in each bin for star-forming galaxies (left panel) and passive galaxies (right panel).  The right-hand axis indicates the black hole mass under the assumption that it is traced by the velocity dispersion. The contours show the underlying density distribution of galaxies, where each outer density contour encloses 90\% of the galaxy population. The arrow angle shows the direction of the steepest metallicity gradient (i.e. following the colour shading) and is calculated (from the horizontal) by the ratio of the partial correlation coefficients between the stellar metallicity, stellar mass and central velocity dispersion. For star-forming galaxies, the left panel shows that both the arrow and the colour-shading follow the stellar mass (x axis) more than the central velocity dispersion (y axis) meaning that the primary driver in increasing the stellar metallicity of star-forming galaxies is the stellar mass. However, for passive galaxies, the right panel shows the opposite, where the stellar metallicity gradient traced by the arrow and colour shading follow more closely the central velocity dispersion (y axis) than the stellar mass (x axis). This confirms the result of the random forest in a more visual way. More importantly, these diagrams provide the direction of this correlation, i.e. for passive galaxies the greater the central velocity dispersion, the greater the stellar metallicity.

We note that in these 3D diagrams we have reduced the dimensionality of the problem by considering the dependence of the stellar metallicity only on stellar mass and central velocity dispersion. However, this implies that the role of additional secondary parameters (SFR, $\phi _g$, M$_H$, $(1+\delta)$) is somehow taken and embedded in the dependencies on stellar mass and central velocity dispersion, and this explains the fact that the partial correlation arrow and colour shading in Figure \ref{fig:hex} does not exactly reproduce the (stronger) relative role of M$_*$ and $\sigma$ that one would expect from the random forest analysis (Fig.\ref{fig:RF}).

\subsection{The $\sigma_c$/Black Hole-Metallicity Relation for passive galaxies}
Fig. \ref{fig:tracks} provides yet another visualisation of these relative dependencies, by showing 2D cuts of the 3D surface shown in Fig. \ref{fig:hex}. Specifically, Fig. \ref{fig:tracks} shows the stellar metallicity versus central velocity dispersion (lower axis) and the equivalent black hole mass (upper axis) binned by tracks of stellar mass for star-forming (upper left) and passive (upper right) galaxies, and stellar metallicity versus stellar mass binned by tracks of central velocity dispersion (lower left and right, same format). The upper figures show that there is a tight correlation between stellar metallicity and central velocity dispersion for passive galaxies, where the change in stellar mass between the tracks barely affects the relationship. On the contrary, for star-forming galaxies the change in stellar mass results in a significant offset of the tracks.
The lower figures show that this is reversed for stellar metallicity versus stellar mass whilst binning in tracks of central velocity dispersion (or black hole mass).
Summarising, for star-forming galaxies at a given central velocity dispersion, the metallicity depends strongly on stellar mass, while at a given stellar mass the stellar metallicity depends little on central velocity dispersion. These dependencies are swapped for passive galaxies: at a given central velocity dispersion the stellar metallicity does not depend on the stellar mass, while at any given stellar mass the stellar metallicity depends strongly on central velocity dispersion.

This result confirms that, for star-forming galaxies, we recover the standard MZR.  However, for passive galaxies, we find strong evidence for a relation between central velocity dispersion and stellar metallicity ($\sigma$ZR).

 We quantify this $\sigma$ZR with the following linear fit in logarithmic quantities: $\rm log( Z/Z_\odot) = [0.541\pm 0.027]\,log(\sigma_c/kms^{-1})\, -[1.107\pm0.06]$.
Note that we have fitted the median bins of the $\sigma$ZR so as to avoid being heavily biased by the overpopulated centre of the contours. As mentioned previously, and as discussed further in the next section, it is likely that this relationship with $\sigma_c$ is actually tracing a relationship with black hole mass; therefore, we also fit a black hole mass metallicity relation (BHZR), using the black hole masses calculated from $\sigma_c$.
This gives $\rm log( Z/Z_\odot) = [0.103\pm0.01]\;log(M_{BH}/M_\odot -8) -0.716 \pm 0.04$.
The best fit to the $\sigma$ZR (and equivalent BZR) is shown in Fig. \ref{fig:best_fit}.

\section{What is $\sigma_c$ tracing - black hole mass or gravitational potential?}

\begin{figure*}
    \centering
    \includegraphics[width=0.95\columnwidth]{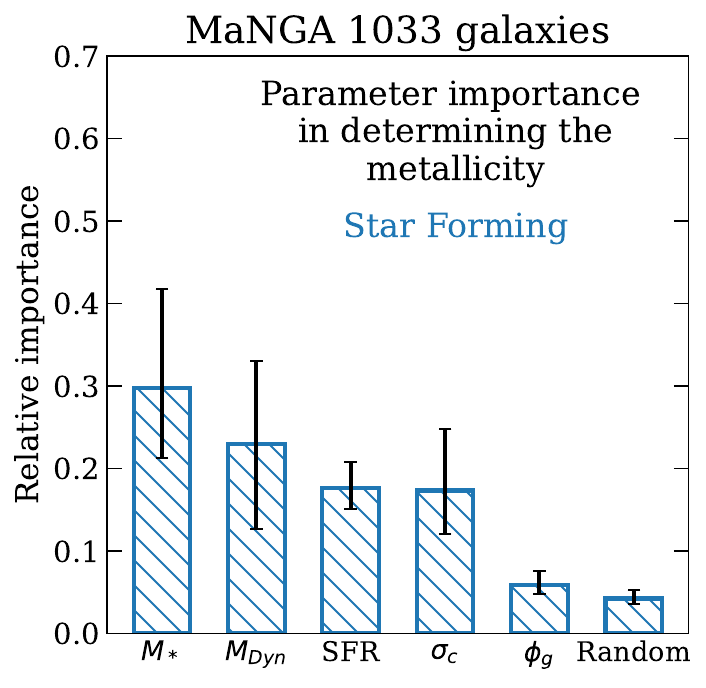}
    \includegraphics[width=0.95\columnwidth]{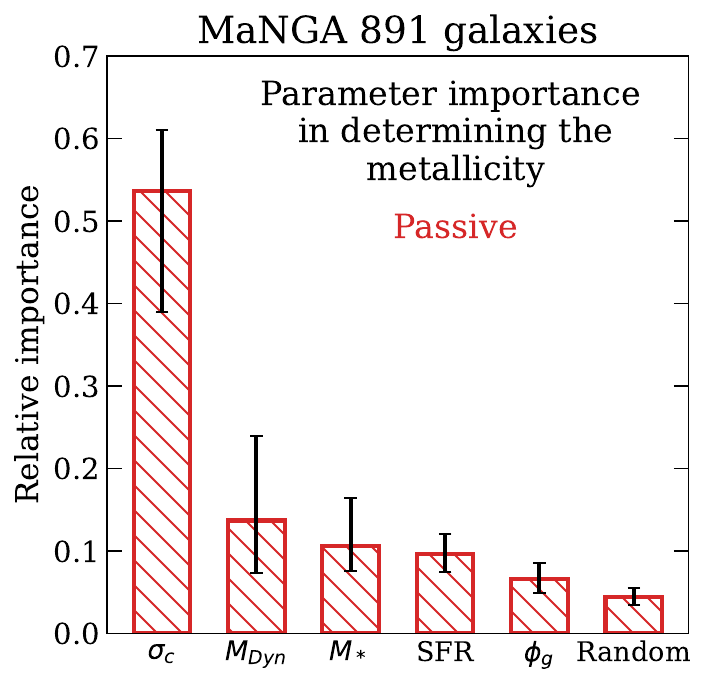}
    \includegraphics[width=0.95\columnwidth]{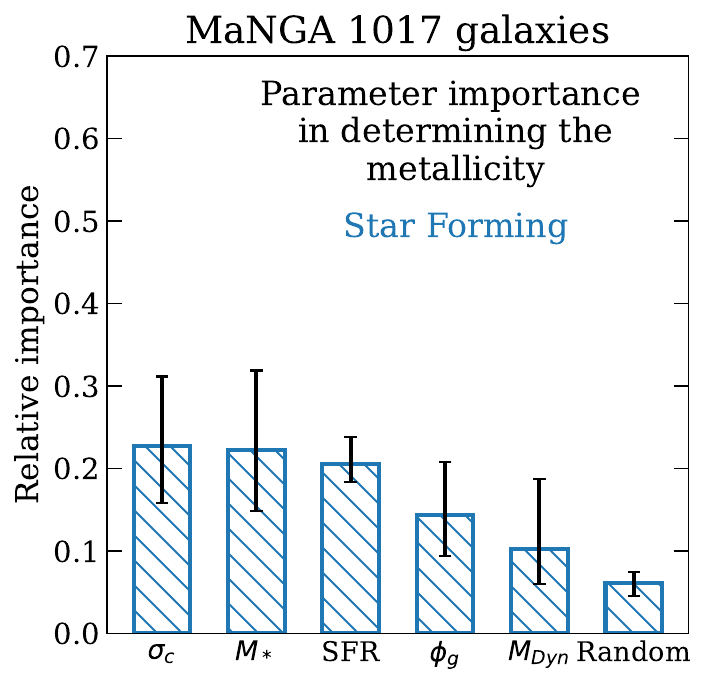}
    \includegraphics[width=0.95\columnwidth]{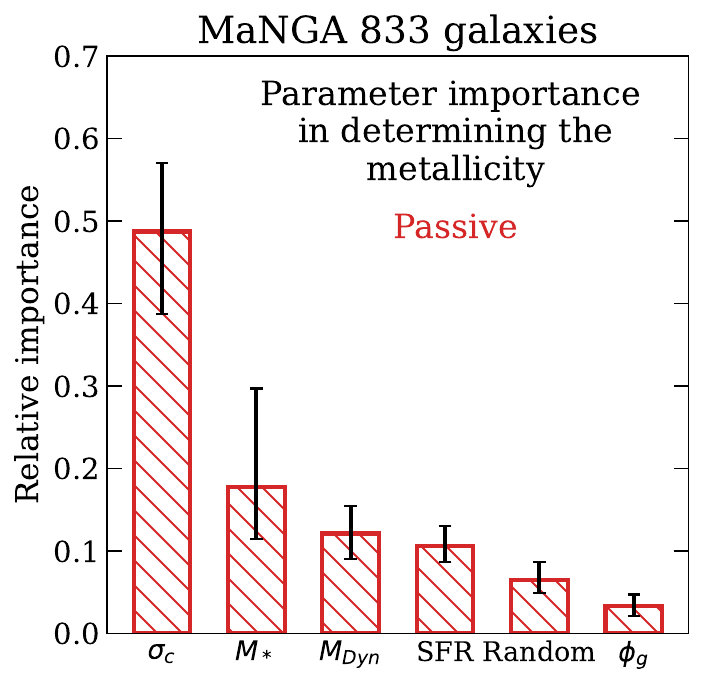}
    \caption{Importance of various galactic properties in determining the stellar metallicity
    for star-forming (left panels) and passive (right panels) galaxies for two samples from the 
    MaNGA survey (\citet{Li2018} top panels, \citet{Brownson2022} bottom panels),
    for which spatially resolved spectroscopy provides accurate dynamical masses and gravitational potentials. Left panels: star-forming galaxies; right panels: passive galaxies. The following galactic parameters are considered: stellar mass ($M_*$),
     central velocity dispersion $\sigma_c$, star formation rate (SFR), the dynamical mass, a control uniform random variable (Random), gravitational potential traced by $\rm \phi_g=M_{dyn}/r_e$ and specific gravitational potential traced by $\rm  \phi_{gs} = V^2 + 3 \sigma^2$. The left-hand panels show that for star-forming galaxies there is no clear primary dependence for the stellar metallicity, due to a combination of a small sample size and difficulty accurately measuring stellar metallicities of star-forming galaxies.
    However, the key result is that for passive galaxies (right-hand panels) the dependence is almost purely on central velocity dispersion, even with the addition of the dynamical mass, and both definitions of the gravitational potential $\phi_g$,  all of which should better trace the gravitational potentials than the central velocity dispersion.
    }
    \label{fig:RF_manga}
\end{figure*}

One important consideration is to properly analyse whether $\sigma_c$ is tracing black hole mass, as suggested by various works \citep{Kormendy2013,Saglia2016,Terrazas2017,Piotrowska2022}, or is in fact tracing the dynamical mass or underlying gravitational potential. We conduct a test of this using $\sim$2000 galaxies from the MaNGA survey with accurate dynamical masses (for more details, see section \ref{sec:manga}). We use two different samples of dynamical measurements. 

The first uses the dynamical masses from \cite{Li2018}, which we also use to define the quantity $\rm \phi_g=M_{dyn}/r_e$ (where $r_e$ is the half-light radius), which is proportional to the gravitational potential. The upper panel of Fig. \ref{fig:RF_manga} shows the random forest regression results for these galaxies. We also use a similar sized sample of kinematic measurements from \cite{Brownson2022}. In this case, the proxy for the (specific) gravitational potential is taken as the quantity $\rm  \phi_{g} = V^2 + 3 \sigma^2$, where V is the rotation velocity and $\sigma$ is the velocity dispersion. 
The lower panel of Fig. \ref{fig:RF_manga} shows the random forest regression results for this sample of galaxies. The results for the two samples appear to agree, at least qualitatively. Essentially, the star-forming sample shows a null result (although the stellar mass is still formally the most predictive quantity for the stellar metallicity), likely due to the small sample size of around ~1000 galaxies, and in particular, the difficulty in measuring stellar metallicities for star-forming galaxies due to their weak stellar continuum and absorption lines. However, the key insight comes from the passive sample. Clearly $\sigma_c$, is the primary driver of the stellar metallicity, even with the addition of dynamical mass and both definitions of $\phi_g$. This shows that the dependence seen on $\sigma_c$ is not a by-product of a dependence on the gravitational potential or with the dynamical mass, rather, we believe it is actually a direct dependence on black hole mass. 

 We note that both the dynamical mass and $\rm  \phi_{g}$ are estimated within $r_e$, likewise the stellar metallicity. Therefore, if $\sigma _c$ was simply tracing dynamical mass and/or $\rm  \phi_{g}$ then the latter two should be predicting the stellar metallicity better, because they are measured within the same region, but this is opposite to what was obtained by the Random Forest analysis.

Summarising, the test performed in this section (although based on lower statistics than the SDSS DR7 sample), disfavours the scenario whereby the stellar metallicity dependence on central velocity dispersion is simply tracing deeper gravitational potentials in more massive galaxies. Rather, this dependence, we believe, is due to the black hole mass itself, resulting from the integrated effect of AGN feedback.

\section{Centrals vs Satellites}
\label{sec:centvssat}
\begin{figure*}
    \centering
    \includegraphics[width=0.99\columnwidth]{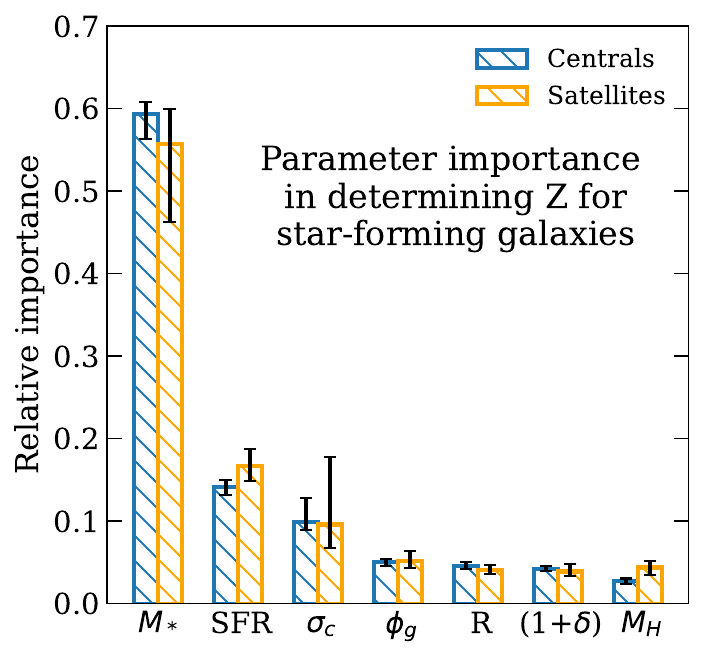}
    \includegraphics[width=0.99\columnwidth]{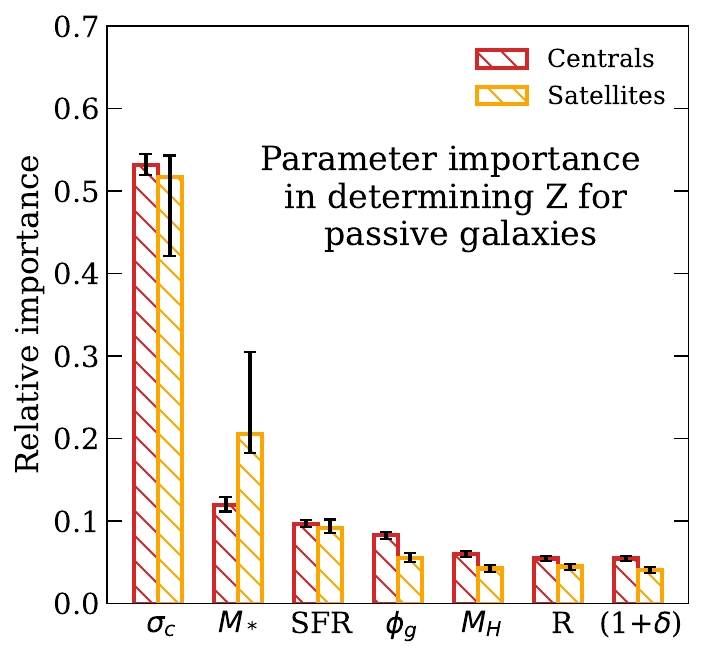}
    \caption{ The parameter importances of various galactic properties in determining the stellar metallicity (Z) for star-forming (left panel) and passive (right panel) galaxies, split into central (blue or red, wider bars) or satellite galaxies (orange, thinner bars) . The galactic parameters evaluated are stellar mass ($M_*$), star formation rate (SFR),  central velocity dispersion ($\sigma_c$), the gravitational potential ($\phi_g$), a control uniform random variable (Random), overdensity (1+$\delta$), and halo mass ($M_H$). We see from both panels that there is little difference between centrals and satellites.}
    \label{fig:RF_cent_sat}
\end{figure*}
We can explore whether these relations that we have found are the same for central and satellite galaxies, and/or there is any environmental effect. This is another test of whether environment plays a significant role. The differences between the stellar metallicities of central and satellite galaxies have been highlighted previously in \cite{Peng2015}. We can probe whether this is the case by using our random forest regression methodology. The first stage is to split the dataset into central and satellite galaxies. We use a mass-weighted approach for this, with central galaxies classified as the most massive galaxy in their group \citep{Yang2005, Yang2007}, with the remaining galaxies being classified as satellites. We also test using a light-weighted classification and find it makes no significant difference.
Fig. \ref{fig:RF_cent_sat} shows the random forest regression parameter importances, in exactly the same form as Fig. \ref{fig:RF}, only this time with the star-forming and passive samples split into central or satellite galaxies based upon whether they are the most massive galaxy in their group. Fig. \ref{fig:RF_cent_sat} illustrates that the results remain unchanged regardless of central or satellite classification, i.e. again we find stellar metallicity is not driven by environmental effects. Interestingly, this is different to what was found in \cite{Peng2015}, which we attribute to being able analyse almost all relevant parameters simultaneously with the random forest, which was not possible in the case of \cite{Peng2015}. 

 On a first glance this may seem surprising - there is plenty of evidence that the quenching of satellite galaxies is linked to environmental effects \citep{Davies2019, Samuel2022}. However, it has been shown recently that there appears to be a mass dependence as to the quenching mechanisms of satellite galaxies \citep[see for example][]{Goubert2024}. Higher-mass satellites would be capable of quenching via integrated AGN feedback, whilst lower-mass satellites would be quenched via environment \citep{Goubert2024}.  Our finding that for the passive satellites in our sample, Z is driven by $\sigma_c$ (or $M_{BH}$) supports this scenario as we are primarily probing the higher-mass satellite population. This is due to our strict S/N requirement for obtaining reliable stellar metallicities.

\section{Summary, discussion and interpretation}

In summary, we have found that in star-forming galaxies the stellar metallicity is driven primarily by the stellar mass in the mass-metallicity relation (MZR). This is likely tracing the integral of metal formation \citep[as seen for gas-phase metallicities in][]{Baker2023c}{}{}: more massive galaxies have produced a greater mass of stars, hence a greater amount of metals. We also find a secondary dependence on SFR consistent with the stellar version of the Fundamental Metallicity Relation (FMR) \citep{Looser23}.
 However, we have discovered that for passive galaxies the stellar metallicity is driven primarily by central stellar velocity dispersion, in a $\sigma$ZR. We have shown that such a relation with $\sigma_c$ is unlikely to be tracing a link with dynamical mass or gravitational potential, and that it is most likely tracing a relation with black hole mass. The black hole mass itself provides an indication of the integrated history of feedback associated with black hole accretion,  so this $\sigma$ZR we believe stems from the black hole quenching galaxies via halo heating (either preventative or delayed feedback), which suppresses gas accretion and results in 'starvation'. Residual star formation during starvation results in a rapid increase in metallicity as a consequence of closed-box evolution without any diluting inflows \citep{Peng2015,Trussler2020}.

This result crucially bridges the previous gap between studies of the stellar metallicities, finding evidence for starvation being a key quenching mechanism, and the causal relation between central velocity dispersion (a strong tracer of black hole mass) and quenching.

\section*{Acknowledgements}
WB, RM, acknowledge support by the Science and Technology Facilities Council (STFC), by the ERC Advanced Grant 695671 ‘QUENCH’, and by the
UKRI Frontier Research grant RISEandFALL. RM also acknowledges funding from a research professorship from the Royal Society. 

Funding for the Sloan Digital Sky Survey IV has been provided by the Alfred P. Sloan Foundation, the U.S. Department of Energy Office of Science, and the Participating Institutions. SDSS acknowledges support and resources from the Center for High-Performance Computing at the University of Utah. The SDSS web site is www.sdss.org.

SDSS is managed by the Astrophysical Research Consortium for the Participating Institutions of the SDSS Collaboration including the Brazilian Participation Group, the Carnegie Institution for Science, Carnegie Mellon University, Center for Astrophysics | Harvard \& Smithsonian (CfA), the Chilean Participation Group, the French Participation Group, Instituto de Astrofísica de Canarias, The Johns Hopkins University, Kavli Institute for the Physics and Mathematics of the Universe (IPMU) / University of Tokyo, the Korean Participation Group, Lawrence Berkeley National Laboratory, Leibniz Institut für Astrophysik Potsdam (AIP), Max-Planck-Institut für Astronomie (MPIA Heidelberg), Max-Planck-Institut für Astrophysik (MPA Garching), Max-Planck-Institut für Extraterrestrische Physik (MPE), National Astronomical Observatories of China, New Mexico State University, New York University, University of Notre Dame, Observatório Nacional / MCTI, The Ohio State University, Pennsylvania State University, Shanghai Astronomical Observatory, United Kingdom Participation Group, Universidad Nacional Autónoma de México, University of Arizona, University of Colorado Boulder, University of Oxford, University of Portsmouth, University of Utah, University of Virginia, University of Washington, University of Wisconsin, Vanderbilt University, and Yale University.


\section*{Data Availability}

The SDSS DR7 data is publicly available. The MaNGA data used is also publicly available at \url{https://www.sdss.org/dr15/manga/manga-data/}. 
The MPA-JHU catalogue is publicly available at \url{https://wwwmpa.mpa-garching.mpg.de/SDSS/DR7/}.



\bibliographystyle{mnras}
\bibliography{refs} 

\appendix

\section{Reliability of $\sigma$ as a tracer of black hole mass}

Another question is exactly how reliable $\sigma_c$ (the central stellar velocity dispersion) is as a tracer of black hole mass. This has previously been explored in \cite{Piotrowska2022}, where they found that out of stellar mass, bulge mass and stellar velocity dispersion it was the stellar velocity dispersion that most tightly correlated with the black hole mass (as obtained from the directly measured black hole masses in \cite{Terrazas2017}). 

\section{Random Forest hyper-parameters}

Table \ref{tab:rf_params} shows the random forest hyper-parameters and the passive galaxy sample mean squared errors of the test and training sample.

\begin{table*}
\caption{Table showing random forest regression hyper-parameters (from left to right), size of the test sample compared to training sample, the number of estimators, the minimum sample size at the end (leaf) node, the maximum depth of the decision trees, and the number of times the forest was bootstrapped for the percentiles.}
\centering
\label{table:hyper}
\begin{tabular}{l c c c c c c c}
\toprule
\multirow{2}{*}{} & \multirow{2}{*}{$\rm test\_size$} & \multirow{2}{*}{$\rm n\_est$}   & \multirow{2}{*}{$\rm min\_samp\_leaf$}   & \multirow{2}{*}{$\rm max\_depth$}  & \multirow{2}{*}{$\rm n\_times$} & \multirow{2}{*}{MSE test} & \multirow{2}{*}{MSE train}\\ \\[+5pt]

\midrule

 SDSS  & 50\%& 200 & 15 & 100  & 100 & 0.007 &  0.005  \\
 MaNGA  & 20\%& 200 & 15 & 100 & 100 & 0.008 & 0.005   \\
\bottomrule
\label{tab:rf_params}
\end{tabular}
\end{table*}


\bsp	
\label{lastpage}
\end{document}